\documentclass{pasj00}

\begin{document}
\SetRunningHead{M.~Sato et al.}{Absolute Proper Motions of H$_2$O Masers in NGC~281}
\Received{2007/02/04}%%{yyyy/mm/dd}
\Accepted{2007/04/18}%%{yyyy/mm/dd}
\SetVolumeData{2007}{59}{4}
\def\pageSTART{??}
%%%\Published{2007/8/25}

\title{Absolute Proper Motions of H$_2$O Masers Away from the Galactic Plane \\Measured with VERA in the "Superbubble" Region NGC~281}
%%% begin:list of authors
\author{
Mayumi \textsc{Sato},\altaffilmark{1,2}
Tomoya \textsc{Hirota},\altaffilmark{1,3}
Mareki \textsc{Honma},\altaffilmark{1,3}
Hideyuki \textsc{Kobayashi},\altaffilmark{1,2,3}
Tetsuo \textsc{Sasao},\altaffilmark{4,5}\\
Takeshi \textsc{Bushimata},\altaffilmark{1,6}
Yoon Kyung \textsc{Choi},\altaffilmark{1,2}
Hiroshi \textsc{Imai},\altaffilmark{7}
Kenzaburo \textsc{Iwadate},\altaffilmark{1}
Takaaki \textsc{Jike},\altaffilmark{1}\\
Seiji \textsc{Kameno},\altaffilmark{7}
Osamu \textsc{Kameya},\altaffilmark{1,3}
Ryuichi \textsc{Kamohara},\altaffilmark{1}
Yukitoshi \textsc{Kan-ya},\altaffilmark{8}
Noriyuki \textsc{Kawaguchi},\altaffilmark{1,3}\\
Masachika \textsc{Kijima},\altaffilmark{3}
Mi Kyoung \textsc{Kim},\altaffilmark{1,2}
Seisuke \textsc{Kuji},\altaffilmark{1}
Tomoharu \textsc{Kurayama},\altaffilmark{1}
Seiji \textsc{Manabe},\altaffilmark{1,3}\\
Kenta \textsc{Maruyama},\altaffilmark{9}
Makoto \textsc{Matsui},\altaffilmark{9}
Naoko \textsc{Matsumoto},\altaffilmark{9}
Takeshi \textsc{Miyaji},\altaffilmark{1}
Takumi \textsc{Nagayama},\altaffilmark{9}\\
Akiharu \textsc{Nakagawa},\altaffilmark{7}
Kayoko \textsc{Nakamura},\altaffilmark{9}
Chung Sik \textsc{Oh},\altaffilmark{1,2}
Toshihiro \textsc{Omodaka},\altaffilmark{7}
Tomoaki \textsc{Oyama},\altaffilmark{1}\\
Satoshi \textsc{Sakai},\altaffilmark{1}
Katsuhisa \textsc{Sato},\altaffilmark{1}
Katsunori M. \textsc{Shibata},\altaffilmark{1,3}
Motonobu \textsc{Shintani},\altaffilmark{9}
Yoshiaki \textsc{Tamura},\altaffilmark{1,3}\\
Miyuki \textsc{Tsushima},\altaffilmark{9}
and Kazuyoshi \textsc{Yamashita}\altaffilmark{3}
}

%%%\altaffiltext{1}{Mizusawa VERA Observatory Mitaka Office, National Astronomical Observatory, 2-21-1 Osawa, Mitaka, Tokyo 181-8588}
\altaffiltext{1}{Mizusawa VERA Observatory, National Astronomical Observatory, 2-12 Hoshi-ga-oka, Mizusawa-ku, Oshu, Iwate 023-0861}
\email{mayumi.sato@nao.ac.jp}
\altaffiltext{2}{Department of Astronomy, Graduate School of Science, The University of Tokyo, 7-3-1 Hongo, Bunkyo-ku, Tokyo 113-0033}
\altaffiltext{3}{Department of Astronomical Sciences, Graduate University for Advanced Studies, 2-21-1 Osawa, Mitaka, Tokyo 181-8588}
\altaffiltext{4}{Department of Space Survey and Information Technology, Ajou University, Suwon, Republic of Korea}
\altaffiltext{5}{Korean VLBI Network, KASI, Seoul, Republic of Korea}   
\altaffiltext{6}{Space VLBI Project, National Astronomical Observatory, 2-21-1 Osawa, Mitaka, Tokyo 181-8588}
\altaffiltext{7}{Faculty of Science, Kagoshima University, 1-21-35 Korimoto, Kagoshima, Kagoshima 890-0065}
\altaffiltext{8}{Department of Astronomy, Yonsei University, Seoul, Republic of Korea}
\altaffiltext{9}{Graduate School of Science and Engineering, Kagoshima University, 1-21-35 Korimoto, Kagoshima, Kagoshima 890-0065}
%%% end:list of authors
%%
%%% Please use the following style in case that sorting by 
%%% affilation is impossible. 
%%
%% \author{%
%%   D-Firstname \textsc{D-Familyname}\altaffilmark{1}
%%   E-Firstname \textsc{E-Familyname}\altaffilmark{1,2}
%%   and
%%   F-Firstname \textsc{F-Familyname}\altaffilmark{2}}
%% \altaffiltext{1}{Address of Institute}
%% \email{ddddd@xxx.xxx.xx.xx}
%% \email{eeeee@xxx.xxx.xx.xx}
%% \altaffiltext{2}{Address of Institute}
%%
%% `\KeyWords{}' always has to be placed before `\maketitle'.
%%\KeyWords{xxxx:xxxx ......} %Do NOT move this preamble from here!
\maketitle

\begin{abstract}
We report on absolute proper-motion measurements of an H$_2$O maser source in the NGC~281~West molecular cloud, which is located $\sim$320 pc above the Galactic plane
 and is associated with an H\emissiontype{I} loop extending from the Galactic plane.
We have conducted multi-epoch phase-referencing observations of the maser source with VERA (VLBI Exploration of Radio Astrometry) over a monitoring period of 6 months since May 2006.
We find that the H$_2$O maser features in NGC~281 West are systematically moving toward the southwest and further away from the Galactic plane with a vertical velocity of $\sim$20$-$30~km~s$^{-1}$ at its estimated distance of 2.2$-$3.5 kpc.
Our new results provide the most direct evidence that the gas in the NGC 281 region on the H\emissiontype{I} loop was blown out from the Galactic plane, most likely in a superbubble driven by multiple or sequential supernova explosions in the Galactic plane.
\end{abstract}

\section{Introduction}

Over the past few decades, numerous efforts have been made both observationally and theoretically to understand the large-scale superstructures such as superbubbles/supershells in the interstellar medium in the Galaxy and in nearby galaxies.
These large-scale superstructures are characterized by their circular or arc-like appearances with radii ranging from tens of parsecs up to 1000 pc.
Many of these Galactic shells were found by Heiles (1979) based on the survey of the H\emissiontype{I} 21-cm line by Weaver and Williams (1973).
The energy required for such a large-scale superstructure with a radius greater than 100 pc ranges up to 10$^{54}$ erg and is hundreds of times larger than that available from a single supernova explosion (Heiles 1979).
For many years, the long-term evolution of supernova remnants (SNRs) has been studied (e.g., Cox 1972; Chevalier 1974; Tomisaka \& Ikeuchi 1986; Hanayama \& Tomisaka 2006), and 
those superbubbles/supershells are considered to have been formed by multiple or sequential supernovae and stellar winds from OB associations (e.g., Tomisaka \etal\ 1981; Tomisaka \& Ikeuchi 1986; Mac~Low \etal\ 1989; Tomisaka 1998).
An excellent review of the field is given by Tenorio-Tagle and Bodenheimer (1988).
Subsequent studies of the superstructures have led to an extended study of the disk-halo interaction in the Galaxy and in other galaxies (Mac~Low \etal\ 1989; Norman \& Ikeuchi 1989) and,
 as represented by the concept of galactic {`}chimneys{'}, the blowout mechanism of superbubbles has drawn attention as a form of efficient energy transport from a galaxy's disk to its halo (e.g., Norman \& Ikeuchi 1989; Normandeau \etal\ 1996).

For studying the origin, energetics and the timescale of superbubbles/supershells, it is of great importance to investigate the kinematics and dynamics of atomic or molecular clouds and H\emissiontype{II} regions associated with these superstructures.
Whereas theoretical simulation models of supernova evolution have made considerable progress and become increasingly more complex, there had been hardly any means to measure the proper motions of superstructure blowouts observationally, due to insufficient astrometric accuracy of available instruments.
However, recent developments in VLBI technology, especially the phase-referencing VLBI technique, provide a powerful tool with its highest possible precision for absolute astrometry to overcome these observational difficulties and offer a breakthrough in our understanding of the superstructure mechanism in the Galaxy.

The NGC~281 (Sh~184; at $\alpha_{2000}=$00$^{\rm h}$52$^{\rm m}$, $\delta_{2000}=+$56$^\circ$\timeform{34'} or $l=$123$^\circ$.07, $b=-$6.$^\circ$31) region is a good observational target for studying the blowout dynamics of superbubbles.
Based on H\emissiontype{I} data from Hartmann and Burton (1997), Megeath \etal\ (2002) and Megeath \etal\ (2003) identified the NGC~281 molecular cloud complex, situated $z\sim$320 pc above the midplane of the Perseus arm of the Galaxy, on a large-scale H\emissiontype{I} loop from the Galactic plane.
They suggested that these molecular clouds at such a remarkable vertical height from the Galactic plane have formed in the gas swept up and compressed in a blowout triggered by multiple supernova explosions.

The NGC~281 nebula is an H\emissiontype{II} region surrounded by a giant molecular cloud complex which was entirely mapped in the CO emission line by Lee and Jung (2003).
The determination of the photometric/kinematic distance to the NGC~281 region has been attempted for many years, but the obtained values vary widely from 2~kpc to 3.5~kpc (e.g., Sharpless 1954; Georgelin \& Georgelin 1976; Roger \& Pedlar 1981; Henning \etal\ 1994; Guetter \& Turner 1997; Lee \& Jung 2003).
The source of ionization for the NGC~281 nebula is a star cluster whose brightest member is a compact, Trapezium-like OB multiple star system called ADS~719 or HD~5005 (Sharpless 1954; Elmegreen \& Lada 1978; Henning \etal\ 1994; Guetter \& Turner 1997).
A southwestward proper motion of HD~5005 was observed with {\it Hipparcos} (Perryman \etal\ 1997), but with insufficient precision to discuss the motion of the system.

Because of the clear separation between the H\emissiontype{II} region and the surrounding molecular clouds, along with its high Galactic latitude with negligible contamination by interstellar matter, the NGC~281 region has been a good laboratory for studying star formation at multiple wavelengths for more than a decade.
Elmegreen and Lada (1978) discovered an H$_2$O maser source at the $^{12}$CO emission peak in the NGC~281 West molecular cloud (located $\sim$\timeform{5'} to the southwest of HD~5005) near the molecular/H\emissiontype{II} interface and proposed that the expansion of the H\emissiontype{II} region is triggering star formation in the adjoining molecular cloud.
At this location of the $^{12}$CO emission peak in the NGC~281 West cloud, there is a far-infrared source IRAS~00494+5617, and an embedded cluster of low mass stars has also been detected in the near-infrared observations (Carpenter \etal\ 1993; Hodapp 1994; Megeath \& Wilson 1997).
Tofani \etal\ (1995) observed the H$_2$O maser source in NGC~281 West cloud with the VLA and found that the H$_2$O maser emission is resolved into three distinct spatial components C1, C2 and C3 with maximum separation of $\sim$\timeform{8''}.

The H$_2$O maser source in the NGC~281 West cloud provides an excellent target for high-resolution VLBI observations.
The average of the maser motions is expected to trace the systemic motion of the molecular cloud on the H\emissiontype{I} loop, which then yields a direct measure of the superbubble expansion.
In order to detect the systemic motion of the NGC 281 West cloud on the H\emissiontype{I} loop, we therefore conducted high-resolution observations of the H$_2$O maser source in the cloud with VERA, a Japanese VLBI array dedicated to phase-referencing VLBI astrometry (e.g., Honma \etal\ 2000; Kobayashi \etal\ 2003; Honma \etal\ 2005; Imai \etal\ 2006).
In this paper, we report on our successful detection of the absolute proper motion of the H$_2$O maser source, which reveals the motion of this region away from the Galactic plane, most likely having originated in the Galactic plane and been blown out by a superbubble blowout.

\section{VERA Observations and Data Reduction}

The VERA observations were conducted at six epochs, spaced approximately at monthly intervals, on 2006 May 14, July 21, August 3, September 5, October 25, and November 18 (day of year 134, 202, 215, 248, 298, and 322, respectively).
In each epoch, the observation was carried out for 7$-$9 hours in the dual-beam mode for phase referencing (e.g., Honma \etal\ 2003).
The calibration data of real-time phase difference between the two beams were taken during the observations, using artificial noise sources injected into two beams at each station (Kawaguchi \etal\ 2000).
An H$_2$O maser source in the NGC~281 West cloud was observed simultaneously with one of the two alternately-scanned position-reference sources
 J0047+5657 (0$^\circ$.84 separation; at $\alpha_{2000}=$00$^{\rm h}$47$^{\rm m}$00.$^{\rm s}$428805, $\delta_{2000}=+$56$^\circ$\timeform{57'}\timeform{42.''39479} by Beasley \etal\ 2002)
 and J0042+5708 (1$^\circ$.50 separation; at $\alpha_{2000}=$00$^{\rm h}$42$^{\rm m}$19.$^{\rm s}$451727, $\delta_{2000}=+$57$^\circ$\timeform{08'}\timeform{36.''58602} by Beasley \etal\ 2002), switched typically every 10 minutes.
A bright calibrator source was also observed every hour at each epoch: J0319+4130 ($=$ 3C~84) for the first epoch and J2232+1143 ($=$ CTA~102) for the last five epochs.
Left-hand circularly polarized signals were digitized at 2-bit sampling and recorded at a data rate of 1024~Mbps. 
Among the total bandwidth of 256~MHz, one 16-MHz IF channel was assigned to the H$_2$O maser source in NGC~281 West and the other 15 IF channels of each 16-MHz bandwidth were assigned to one of the position-reference sources with the VERA digital filter unit (Iguchi \etal\ 2005).
The data correlation was performed with the Mitaka FX correlator (Chikada \etal\ 1991).
In order to achieve a higher spectral resolution for the H$_2$O maser lines, only the central 8~MHz of the total 16-MHz bandwidth of the IF channel for the maser lines was split into 512 spectral points, yielding frequency and velocity resolutions of 15.625~kHz and 0.21~km~s$^{-1}$, respectively.
The observed frequencies of the maser lines were converted to the radial velocities with respect to the local standard of rest (LSR), $V_{\rm LSR}$, using a rest frequency of 22.235080~GHz for the H$_2$O 6$_{16}$-5$_{23}$ transition.
The system noise temperatures at the zenith were typically 150$-$300~K for all epochs except for the second and fourth epochs, where the system temperatures were 400$-$500~K. 
The aperture efficiencies of the antennas ranged from 45\% to 52\%.

Calibration and imaging were performed in a standard manner with the NRAO Astronomical Image Processing System (AIPS) package.
For single-beam data analyses to discuss the internal motion of the H$_2$O maser source in NGC~281 West, the visibilities of all velocity channels at each epoch were phase-referenced to the brightest reference maser feature at $V_{\rm LSR}=-$31.8~km~s$^{-1}$ (feature 4 in table 1).
After those calibration processes, spectral-line image cubes were made using the AIPS task IMAGR.
In order to search for maser features that were far from the reference feature, we first made wide-field maps of 1024 pixels $\times$ 1024 pixels with a pixel size of 1 mas to obtain a field of view of $\sim$\timeform{1''}, locating the map centers at expected positions of the maser components C1, C2 and C3 (Tofani \etal\ 1995).
Around the detected maser features, narrower-field maps were then made with 512 pixels $\times$ 512 pixels of size 0.05 mas, yielding a field of view of 25.6~mas $\times$ 25.6~mas, and the positions of these features were determined with respect to the reference maser feature.
The synthesized beam has an FWHM beam size of 1.3~mas $\times$ 0.8~mas with a position angle of $-$43$^\circ$.
RMS noise levels per channel were 50$-$150~mJy.

Based on our criteria for detection, all maser features presented here were detected with a signal-to-noise ratio higher than 7 over two or more velocity channels.
In order to avoid the effect of maser structure on position measurements, each maser spot was averaged over the detected velocity channels with the AIPS task SUMIM
 to yield one maser feature, and the feature positions were then determined with respect to the reference feature by the peaks of Gaussian-model fitting with the AIPS task JMFIT.
For single-beam multi-epoch data analyses, we identified maser features at different epochs as being the {`}same{'} on the basis that their positions should agree with those expected from the proper motions estimated at three or more epochs with an accuracy of 1~mas
 and that their radial velocities must be within twice the velocity resolution, 0.42~km~s$^{-1}$, from the mean value for all six epochs.

For dual-beam phase-referencing data analyses to discuss the absolute positions and absolute proper motions of the H$_2$O maser source, the reduction was done in the same manner as the single-beam data reduction described above, but using the extragalactic position-reference quasars, J0047+5657 and J0042+5708, as phase-reference calibrators instead of the brightest maser feature in the single-beam case.
In order to correct for instrumental phase errors, the dual-beam phase calibration data, taken with the artificial noise source in each beam, were also included in the dual-beam data reduction.
The maps obtained from the dual-beam data had low dynamic range compared to those of the single-beam data, mainly due to residuals of tropospheric zenith delay, and thus only three relatively bright maser features (features~4, 9 and 10 in table 1) were detected at three or more epochs in the dual-beam data reduction.

\section{Results}
\begin{figure}
  \begin{center}
    \FigureFile(80mm,120mm){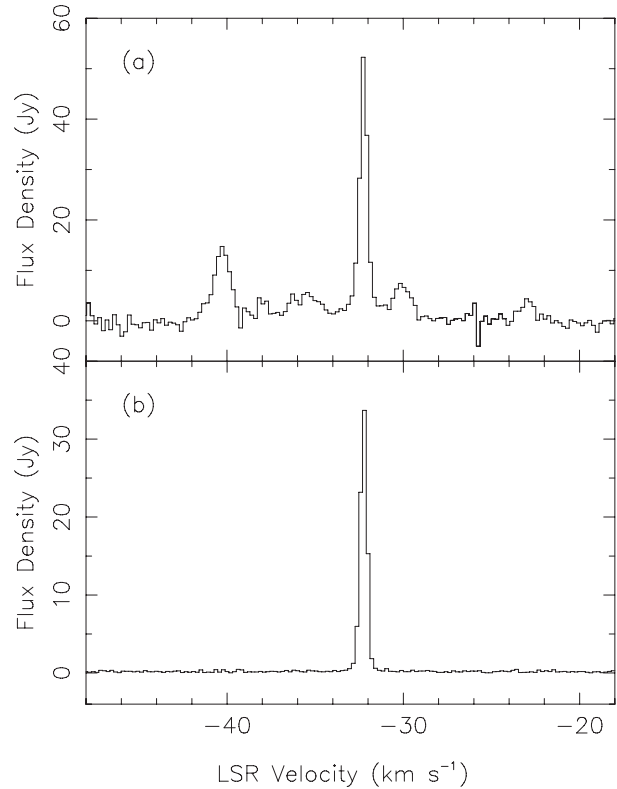}
    %%% \FigureFile(width,height){filename}
  \end{center}
  \caption{The observed spectra of the H$_2$O maser source in NGC~281~West, obtained from a vector average of the received power at the sixth epoch, integrated over the net four-hour observation time: (a) the total power spectrum at Mizusawa station; (b) the cross power spectrum in Mizusawa-Ogasawara baseline.}\label{fig:1}
\end{figure}

\subsection{Feature Distribution and Internal Motions}
First, we show the results from single-beam data analyses to obtain relative positions and internal motions of H$_2$O maser features in NGC~281 West. 
Figures 1a and 1b show examples of total and cross power spectra (autocorrelation and cross-correlation), respectively.
Each spectrum was obtained from a vector average of the received power, integrated over the total observation time at the sixth epoch (approximately net 4 hours). 
The peak intensity occurs at $V_{\rm LSR}\sim -$32~km~s$^{-1}$, which is mainly due to the reference maser feature.
\begin{table*}
\footnotesize
{\bf Table~1.} \hspace{4pt} Results of proper motion measurements.\\
\begin{tabular}{cccrrllrrr}
\hline
 ID & $V_{\rm LSR}$ & Coordinate & $x$~~\, & $y$~~\, & ~~~~~$\mu_X$  & ~~~~~$\mu_Y$  & $\mu_x$~~~~~ & $\mu_y$~~~~~ & Flux~ \\
    & [km s$^{-1}$] & (J2000.0) & [mas] & [mas] & [mas yr$^{-1}$] & [mas yr$^{-1}$] & [mas yr$^{-1}$] & [mas yr$^{-1}$] & \\
 (1)  &  (2)          & (3)    & (4)~\,  & (5)~\,  & ~~~~~(6)  & ~~~~~(7)  & (8)~~~~~ & (9)~~~~~ &(10)~       \\ 
\hline
 (C3)\\
%%%Feature 4  
 1  & $-$34.1 & $\cdots$ &  222.50 &  $-$55.59 & ~~~~~~~~$\cdots$ & ~~~~~~~~$\cdots$ &    1.76 (0.03) & $-$0.29 (0.04) & 3.9\\
%%%Feature 1c
 2  & $-$32.5 & $\cdots$ & $-$3.39 &      1.37 & ~~~~~~~~$\cdots$ & ~~~~~~~~$\cdots$ &    0.72 (0.14) &    0.09 (0.16) & 1.2\\
%%%Feature 1b 
 3  & $-$32.3 & $\cdots$ & $-$5.22 &      2.35 & ~~~~~~~~$\cdots$ & ~~~~~~~~$\cdots$ &    0.30 (0.10) & $-$0.07 (0.21) & 1.2\\
%%%Feature 1a
 4  & $-$31.8 & 00$^{\rm h}$52$^{\rm m}$24.$^{\rm s}$70086 & 0.00 & 0.00 & $-$2.87 (0.26)$^{*1}$ & $-$2.78 (0.37)$^{*1}$ & 0.00~~~~~~~~\, & 0.00~~~~~~~~\, & 39.8\\
    &         & $+$56$^\circ$\timeform{33' 50.''5270}      &      &      & $-$2.92 (0.23)$^{*2}$ & $-$2.54 (0.35)$^{*2}$ &                &                &\\
%%%Feature 8
 5  & $-$29.8 & $\cdots$ &  137.32 & $-$138.43 & ~~~~~~~~$\cdots$ & ~~~~~~~~$\cdots$ & $-$1.16 (0.21) &    0.84 (0.09) & 3.5\\
%%%Feature 2
 6  & $-$29.6 & $\cdots$ &  101.64 &  $-$89.17 & ~~~~~~~~$\cdots$ & ~~~~~~~~$\cdots$ &    0.47 (0.04) & $-$0.41 (0.06) & 0.6\\
%%%Feature 5a
 7  & $-$24.3 & $\cdots$ &  83.62  &  $-$14.14 & ~~~~~~~~$\cdots$ & ~~~~~~~~$\cdots$ &    0.29 (0.07) &    1.48 (0.20) & 0.8\\
%%%Feature 5b
 8  & $-$22.3 & $\cdots$ &  92.21  &  $-$20.94 & ~~~~~~~~$\cdots$ & ~~~~~~~~$\cdots$ &    1.33 (0.16) &    1.72 (0.20) & 3.9\\
\hline
 (C1)\\
%%%Feature 6 
 9  & $-$39.3 & 00$^{\rm h}$52$^{\rm m}$24.$^{\rm s}$17744 & $\cdots$~~ & $\cdots$~~ & $-$4.88 (0.11)$^{*1}$ & $-$1.71 (0.46)$^{*1}$ & $\cdots$~~~~~~ & $\cdots$~~~~~~ & 5.8\\
    &         & $+$56$^\circ$\timeform{33' 43.''2553}      &            &            & $-$4.76 (0.28)$^{*2}$ & $-$1.44 (0.41)$^{*2}$ &                &                &\\
%%%Feature 7
10  & $-$32.0 & 00$^{\rm h}$52$^{\rm m}$24.$^{\rm s}$19986 & $\cdots$~~ & $\cdots$~~ & $-$2.03 (0.23)$^{*1}$ & $-$4.06 (0.79)$^{*1}$ & $\cdots$~~~~~~ & $\cdots$~~~~~~ & 1.1\\
    &         & $+$56$^\circ$\timeform{33' 43.''3907}      &            &            & $-$1.91 (0.37)$^{*2}$ & $-$4.12 (0.69)$^{*2}$ &                &                &\\
\hline
4w  &         &          &         &          & $-$2.89 (0.18)        & $-$2.65 (0.26)        &             &             & \\
C3m &         &          &         &          & $-$2.43 (0.38)        & $-$2.23 (0.40)        & 0.46 (0.33) & 0.42 (0.31) & \\
9w  &         &          &         &          & $-$4.86 (0.10)        & $-$1.56 (0.31)        &             &             & \\
10w &         &          &         &          & $-$1.99 (0.20)        & $-$4.09 (0.52)        &             &             & \\
C1m &         &          &         &          & $-$3.43               & $-$2.83               &             &             & \\
\hline\\
\end{tabular}
\\
(1) The maser feature ID number. See text for details.
(2) The radial LSR velocity in km~s$^{-1}$ averaged for detected epochs. 
(3) The coordinate RA(J2000.0) and DEC(J2000.0) of the absolute position at the first epoch.
(4) (5) The best-fit angular offsets in mas at the first epoch from the reference maser feature 4, toward the east (RA) and north (DEC) (extrapolated for those not detected at the first epoch).
(6)(7) The best-fit absolute proper motions in mas~yr$^{-1}$ (with the associated uncertainties in parentheses), obtained using the position-reference sources $*1$: J0047+5657 and $*2$: J0042+5708.
(8)(9) The best-fit relative proper motions in the component C3 with respect to the reference maser feature 4 (with the associated uncertainties in parentheses).
(10) Peak flux in Jy~beam$^{-1}$ at the sixth epoch.
\end{table*}
\begin{figure*}
  \begin{center}
    \FigureFile(175mm,90mm){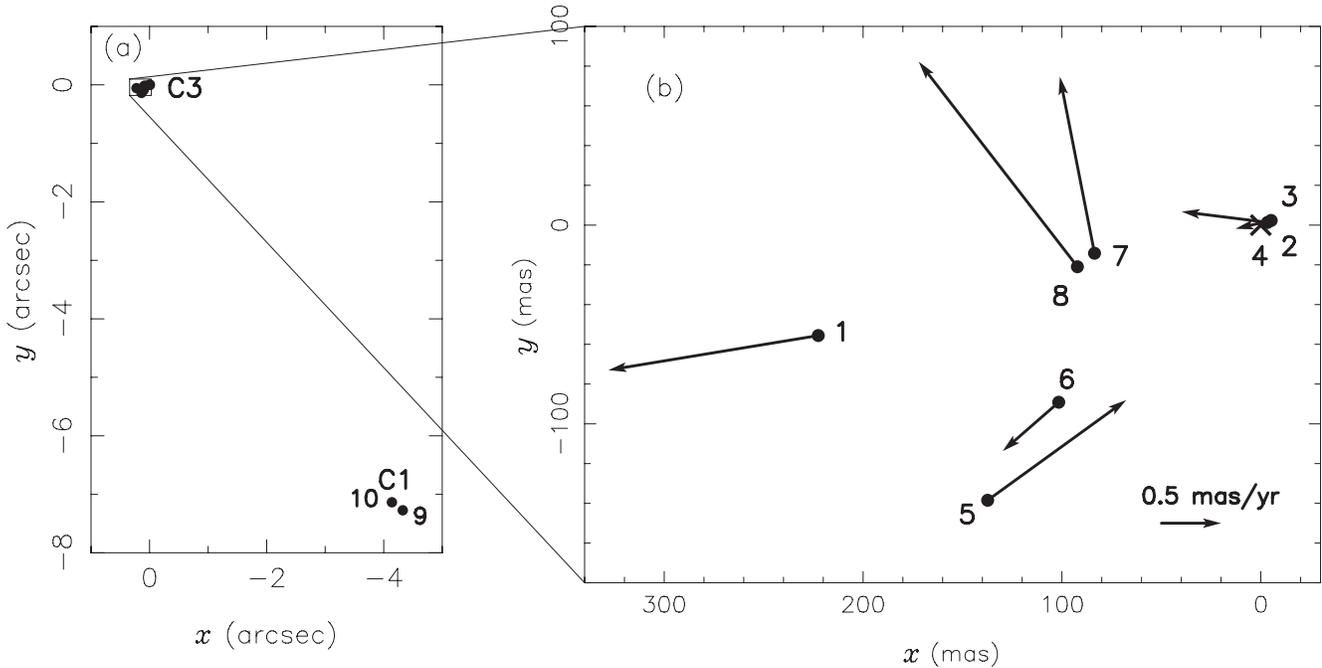}
    %%% \FigureFile(width,height){filename}
    \caption{The maps of the maser feature distribution and internal motions: (a) the maser feature distribution, corresponding well with the relative positions of two maser components C3 and C1 observed with the VLA by Tofani \etal\ (1995); (b) the internal motions of features in C3 relative to feature 4, denoted by X at the map origin (see text). }\label{fig:2}
\end{center}
\end{figure*}
\begin{figure*}
  \begin{center}
    \FigureFile(120mm,90mm){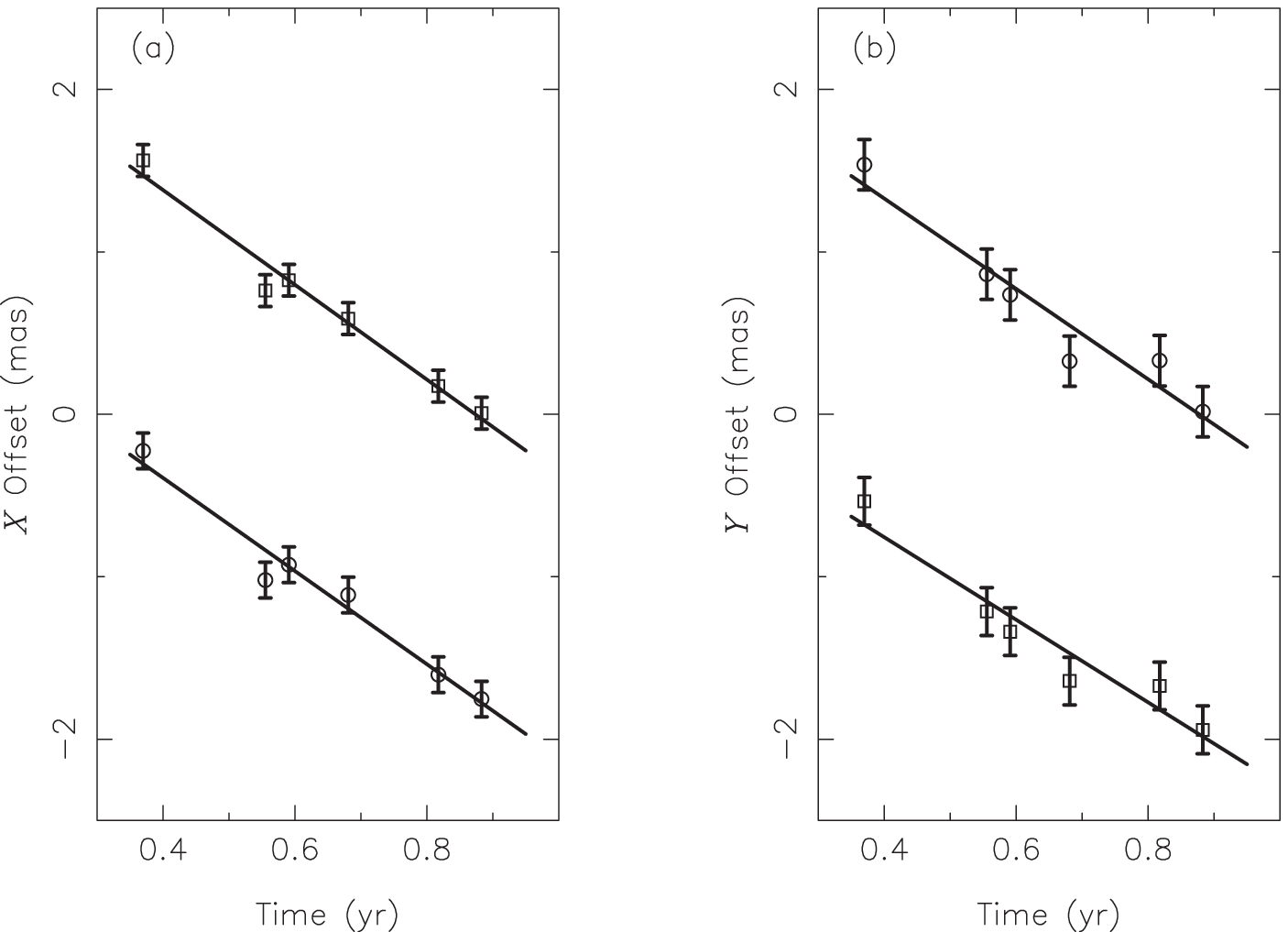}
    %%% \FigureFile(width,height){filename}
  \end{center}
  \caption{Absolute proper-motion measurements of feature~4; using J0047+5657 (open circles) and J0042+5708 (open squares) as position-reference sources. Note that additional offsets are given for clarity: $+$1.0 mas to the squares in (a) and $+$1.5 mas to the circles in (b). The error bars are plotted for the standard deviation $\sigma$ from each linear least-squares fit (see text).}\label{fig:3}
\end{figure*}

Table 1 summarizes the results of both relative and absolute proper-motion measurements obtained by linear least-squares fits.
As listed in table~1, we detected in total 10 maser features with radial velocities $V_{\rm LSR}$ ranging from $-$39 to $-$22~km~s$^{-1}$ at more than two epochs.
The central velocity of these features is in good agreement with the systemic radial velocity obtained from the $^{12}$CO ($J=1-0$) line emission, $V_{\rm LSR}=-31$ km~s$^{-1}$ (Lee \& Jung 2003).
Figure~2a shows the distribution of these maser features, which corresponds well with the relative positions of two maser components C3 and C1 observed with the VLA by Tofani \etal\ (1995).
We did not detect any maser emission at the expected position of the component C2.
With the high spatial resolution of VERA, $\sim$1.2~mas at 22~GHz, our VLBI map separates the H$_2$O maser emission within component C3 into 8 maser features and within C1 into 2 features.

In single-beam data analyses, we measured the accurate relative positions and relative motions of the maser features within C3 only, with respect to the brightest reference feature~4 in C3 ($V_{\rm LSR}=-$31.8~km~s$^{-1}$).
Note that the other two features located within C1 were also used for astrometric measurements in the dual-beam phase-referencing data analyses, which will be discussed in the next section.

The positions of maser features 1$-$3 and 5$-$8 in C3 were determined at each epoch relative to the reference feature 4, and their relative proper motions were also determined at detected epochs by linear least-squares fitting. 
The terms $x$ and $y$ in table~1 are the best-fit angular offsets in mas at the first epoch from the reference feature 4, toward the east (RA) and north (DEC), respectively (extrapolated from the fits for features~5 and 8, which were not detected at the first epoch).
In table~1, $\mu_x$ and $\mu_y$ show the measured relative proper motions with respect to the reference feature~4.
The associated uncertainties in $\mu_x$ and $\mu_y$, as indicated in parentheses in table~1, were estimated from RMS residuals in $x$ and $y$ of 0.1$-$0.8~mas from the linear least-squares fits.

Figure~2b gives a map of the internal motions of the 8 maser features in C3, obtained from the best-fit offsets ($x$,~$y$) at the first epoch with vectors ($\mu_x$,~$\mu_y$) of the relative proper motions.
The reference feature is located at the map origin and denoted by X. 
The best-fit relative positions of the features are denoted by filled circles with the feature ID numbers in table~1. 
Note that the magnitude of the vector ($\mu_x$,~$\mu_y$) is magnified to be the value for 60~yr.
A proper motion of 0.5~mas~yr$^{-1}$ is indicated for reference at the bottom right corner of figure 2b.
This motion, 0.5~mas~yr$^{-1}$, corresponds to 6.9~km~s$^{-1}$ at an estimated distance to NGC~281 of 2.9~kpc (Guetter \& Turner 1997). 
This map indicates expanding motions of the H$_2$O maser features in C3 with velocities from 10 to 20~km~s$^{-1}$.
As given as C3m in table~1, the unweighted mean of the relative proper motions of all 8 maser features in C3 (including feature~4) is ($\bar{\mu_x}$,~$\bar{\mu_y}$)$_{\rm C3}=$(0.46$\pm$0.33,~0.42$\pm$0.31) mas~yr$^{-1}$.  
The magnitude of the mean relative proper-motion vector is 0.62~mas~yr$^{-1}$, which corresponds to 8.5~km~s$^{-1}$ at a distance of 2.9~kpc.
The uncertainties associated with the mean were estimated from the standard deviation of the relative proper motions, divided by $\sqrt {n-1}$, where $n=8$ is the number of the maser features in C3.
The standard deviation of these relative proper motions is (0.88~,0.82) mas~yr$^{-1}$, which corresponds to a deviation of $\sim$10~km~s$^{-1}$ from the mean motion.
\begin{figure}
  \begin{center}
    \FigureFile(79mm,85mm){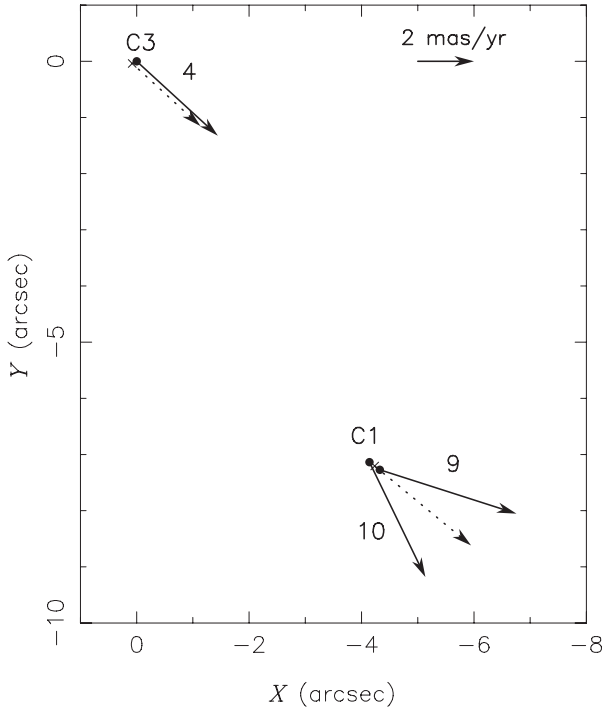}
    %%% \FigureFile(width,height){filename}
  \end{center}
  \caption{A map of the absolute proper motions of the maser features, indicating the measured proper motions of features~4, 9 and 10 (solid arrows) and the mean proper motions of the maser features in C1 and C3 (dotted arrows). The map origin is at RA(J2000.0)$=$00$^{\rm h}$52$^{\rm m}$24.$^{\rm s}$70086 and DEC(J2000.0)$=+$56$^\circ$\timeform{33' 50.''5270} (see text).} \label{fig:4}
\end{figure}

\subsection{Absolute Positions and Absolute Proper Motions}
We now show the results from dual-beam phase-referencing data analyses to obtain absolute positions and proper motions of the H$_2$O maser features in NGC~281 West. 
As shown in table~1, we measured the absolute positions and absolute proper motions $\mu_X$ and $\mu_Y$ toward the east (RA) and north (DEC), respectively, for three bright maser features.
These measurements were performed with respect to two different position-reference sources J0047+5657 and J0042+5708, independently.
For the absolute position of each feature, the unweighted mean of the two values obtained from these two position-reference sources at the first epoch is shown in table~1 (column~3).
The positional difference between the two reference cases was $\sim$0.5 mas, likely due to the uncertainties in the absolute positions of the reference sources: 0.64 mas for J0047+5657 and 0.82 mas for J0042+5708 (Beasley \etal\ 2002). 
These systematic errors do not affect the precision of the proper motion measurements because they only shift the maser feature positions with constant offsets.

Figure~3 shows the absolute proper motion of feature~4.
The open circles and squares indicate values obtained using J0047+5657 and J0042+5708 as position-reference sources, respectively. 
Note, however, that additional offsets are given in figure 3 for clarity: $+$1.0 mas to the squares (from J0042+5708) in figure~3a; and $+$1.5 mas to the circles (from J0047+5657) in figure~3b.
The observation period of 189 days was not long enough to detect the annual parallax significantly.
Hereafter we assume the distance to NGC~281 to be 2.9 kpc (Guetter \& Turner 1997) unless otherwise stated in this paper.
In figures~3a and 3b, the annual parallax at a distance of 2.9 kpc was subtracted from the measured absolute position at each epoch.
The coordinate of the map origin of figures~3 is set to be RA(J2000.0)$=$00$^{\rm h}$52$^{\rm m}$24.$^{\rm s}$70081 and DEC(J2000.0)$=+$56$^\circ$\timeform{33' 50.''5274}.
In figure~3, the error bars are plotted for the standard deviation $\sigma$ from each linear least-squares fit.
Thermal errors in individual position measurements with the AIPS task JMFIT ranged from 0.010~mas to 0.017~mas (well within the size of each circle or square in figure~3), indicating that thermal errors do not dominate or explain enough the deviation of each measurement from the linear fit.
We consider that the main cause of the deviation from the fit is residuals of tropospheric zenith delay, which are difficult to measure quantitatively.
We thus estimate the errors from the standard deviations from the fits. 

The absolute proper motion ($\mu_X$,~$\mu_Y$) of feature~4 was measured to be ($-$2.87$\pm$0.26,~$-$2.78$\pm$0.37) mas~yr$^{-1}$ using J0047+5657 (denoted by $*$1) and ($-$2.92$\pm$0.23,~$-$2.54$\pm$0.35) mas~yr$^{-1}$ using J0042+5708 (denoted by $*$2).
The associated uncertainties in $\mu_X$ and $\mu_Y$, indicated in parentheses in table 1, were estimated from RMS residuals from the linear least-squares fits to the $X$ and $Y$ offsets, respectively.
These RMS residuals in $X$ and $Y$ were 0.10~mas and 0.15~mas, respectively.
The error-weighted mean of these two values ($*$1 and $*$2) is also given in table 1, denoted by 4w, as ($\mu_X$,~$\mu_Y$)$=$($-$2.89$\pm$0.18,~$-$2.65$\pm$0.26) mas~yr$^{-1}$. 

The absolute proper motions $\mu_X$ and $\mu_Y$ of the two features in C1 (features~9 and 10) were also measured in exactly the same manner and are given in table~1.
The error-weighted mean proper motion of each of these features in C1 is, as in table 1, ($\mu_X$,~$\mu_Y$)$=$($-$4.86$\pm$0.10,~$-$1.56$\pm$0.31) mas~yr$^{-1}$ for feature 9 and ($\mu_X$,~$\mu_Y$)$=$($-$1.99$\pm$0.20,~$-$4.09$\pm$0.52) mas~yr$^{-1}$ for feature~10 (denoted by 9w and 10w in table~1, respectively).

Figure~4 gives a vector map of the absolute proper motions 4w, 9w and 10w (shown as solid arrows) measured for features~4, 9 and 10, respectively.
The map origin is set to be the absolute position of feature~4 obtained at the first epoch: RA(J2000.0)$=$00$^{\rm h}$52$^{\rm m}$24.$^{\rm s}$70086 and DEC(J2000.0)$=+$56$^\circ$\timeform{33' 50.''5270}.
Note that the magnitudes of all vectors ($\mu_X$,~$\mu_Y$) are magnified to be the value for 500~yr.
A proper motion of 2~mas~yr$^{-1}$, which corresponds to 27~km~s$^{-1}$ at a distance of 2.9~kpc, is indicated for reference at the top right corner of figure~4. 

In order to obtain the systemic absolute proper motion in C3, we add to the absolute proper motion of feature~4 the unweighted-mean vector of the relative proper motions of the spots in C3 with respect to feature~4, which was obtained in the previous section to be
 ($\bar{\mu_x}$, $\bar{\mu_y}$)$=$(0.46$\pm$0.33,~0.42$\pm$0.31) mas~yr$^{-1}$.
This addition gives, as denoted by C3m in table 1, the systemic absolute proper motion in C3 as ($\bar{\mu_X}$,~$\bar{\mu_Y}$)$_{\rm C3}=$($-$2.43$\pm$0.38,~$-$2.23$\pm$0.40) mas~yr$^{-1}$.
The associated uncertainties were derived from root-sum-square calculations of the errors in the mean proper motion of C3 (relative to feature~4) and in the absolute proper motion of feature~4.
This mean motion is plotted in figure~4 as a dotted arrow from the mean position of the features within C3.

\begin{figure*}
  \begin{center}
    \FigureFile(130mm,60mm){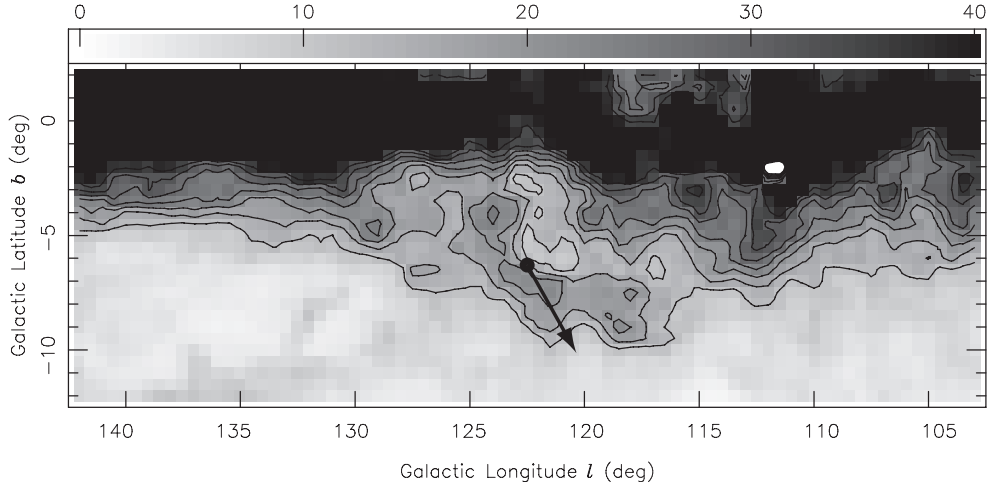}
    %%% \FigureFile(width,height){filename}
  \end{center}
  \caption{A plot of the mean motion of H$_2$O maser features in C3 and C1 relative to the simulated Galactic rotation, superimposed on gray-scale and contour maps of the H\emissiontype{I} data from Hartmann and Burton (1997).  The H\emissiontype{I} maps were obtained by integrating over $V_{\rm LSR}=-65$ to $-$25~km~s$^{-1}$.  The contour peak flux is 75.6~K and the contour levels are 12, 16, 20, 24, 28, 32, 36~K.}\label{fig:5}
\end{figure*}

Since the two features in C1 are measured to be moving away from each other, as seen in figure 4, we assume an expanding internal motion within C1, which is the most common case of jets or outflows from young stellar objects traced by H$_2$O maser motions (e.g., Torrelles \etal\ 2005).
We obtain the unweighted mean of the absolute motions of these features, as denoted by C1m in table 1, to be ($\bar{\mu_X}$,~$\bar{\mu_Y}$)$_{\rm C1}=$($-$3.43,~$-$2.83) mas~yr$^{-1}$.
This mean motion is plotted in figure~4 as a dotted arrow from the mean position of features 9 and 10 in C1.

If we adopt an estimated distance of 2.2~kpc (Georgelin \& Georgelin 1976; Roger \& Pedlar 1981) or 3.5~kpc (Henning \etal\ 1994) instead of 2.9~kpc in the parallax subtraction, the resultant mean proper motions of C3 and C1 vary only by 10$-$15\%.

As seen in figure~4, we find that the maser features in the two components C3 and C1 are systematically moving to the southwest.
This direction is in agreement with that of the proper motion of the multiple star system HD~5005 (the ionizing source of the H\emissiontype{II} region NGC~281), which was measured with {\it Hipparcos} to be ($\mu_X$,~$\mu_Y$)$_{\rm HD5005}=$($-$2.95$\pm$1.37,~$-$3.22$\pm$1.00) mas~yr$^{-1}$ (Perryman \etal\ 1997).
The mean proper motion of maser features in C3 is ($\bar{\mu_X}$,~$\bar{\mu_Y}$)$_{\rm C3}=$($-$2.43$\pm$0.38,~$-$2.23$\pm$0.40) mas~yr$^{-1}$, which agrees with the proper motion of HD~5005 within the margin of error of {\it Hipparcos} measurements.
While the proper motion of HD~5005 measured with {\it Hipparcos} over a few years is associated with a large uncertainty of 30$-$50\%, we successfully obtained the absolute proper motions of the H$_2$O maser features in NGC~281 West with a much better precision over a period of half a year.

\section{Discussion}

\subsection{Vertical Motion Away from the Galactic Plane}

With the results of our proper-motion measurements, we now consider the motion of the H$_2$O maser source in NGC~281 West with respect to the Galactic plane.
At the position of NGC~281 West, the Galactic plane lies almost parallel to the east-west (RA) direction with a position angle of 90.$^\circ$2,
 so that we can regard the motions of the maser source in the directions of RA and in DEC as those parallel and perpendicular to the Galactic plane, respectively.

In this section, we do not discuss in detail the RA motions parallel to the Galactic plane because of their relatively large uncertainties due to the uncertainties in the Galactic rotation. 
We therefore focus on the motion of the maser source in NGC~281 perpendicular to the Galactic plane, called the {\it vertical} motion hereafter, represented by the DEC motion.

If we assume the distance from the Sun to the Galactic center, $R_0$, to be 8.0~kpc (Reid 1993), the LSR velocity $\Theta_0$ of the Galactic rotation at $R_0$ to be 200~km~s$^{-1}$ (e.g., Kalirai \etal\ 2004; Avedisova 2005), and a flat rotation curve of the Galaxy,
 then the expected apparent motion of the NGC~281 region due to the Galactic rotation seen at the LSR is ($\mu_{{\rm G}X}$,~$\mu_{{\rm G}Y}$)$=$($-$2.75,~$-$0.25) mas~yr$^{-1}$ in RA and DEC, including the apparent motion in DEC (in $b$) arising from the nonzero galactic latitude of NGC~281 ($b=-$6.$^\circ$31).
Here the expected radial velocity of the NGC~281 at its position (calculated from the flat rotation model) is $V_{\rm LSR}=-31.8$~km~s$^{-1}$, which agrees well with the observed radial velocities of the NGC~281~West cloud and the H$_2$O maser features in the cloud.  
If we adopt the solar motion relative to the LSR based on {\it Hipparcos} data, corresponding to a solar motion toward the coordinates $\alpha_{2000}=$251.$^\circ$51, $\delta_{2000}=$10.$^\circ$13 with the velocity $V=$13.34~km~s$^{-1}$ (Dehnen \& Binney 1998), we also obtain the solar-motion effect to the observed motion of the NGC~281 region to be ($\mu_{{\rm S}X}$,~$\mu_{{\rm S}Y}$)$=$(0.81,~$-$0.51) mas~yr$^{-1}$ in RA and DEC.
Subtracting these apparent motions due to the Galactic rotation and the solar motion from the measured absolute proper motion of the H$_2$O maser source in NGC~281 West, we therefore obtain the motion of the maser source with respect to the rotation of the Galactic Plane.
The mean motions of C3 and C1 with respect to the Galactic rotation are obtained from the results in the previous section to be ($\mu_X$,~$\mu_Y$)$_{{\rm C3}-{\rm GR}}=$($-$0.49,~$-$1.47) mas~yr$^{-1}$ and ($\mu_X$,~$\mu_Y$)$_{{\rm C1}-{\rm GR}}=$($-$1.49,~$-$2.07) mas~yr$^{-1}$, respectively.
The mean of these two motions is then ($\mu_X$,~$\mu_Y$)$_{{\rm sys}-{\rm GR}}=$($-$0.99,~$-$1.77)~mas~yr$^{-1}$, which is considered to trace the systemic motion with respect to the Galactic rotation.
Figure 5 depicts this mean motion with respect to the simulated Galactic rotation, superimposed on grayscale and contour maps based on the H\emissiontype{I} data from Hartmann and Burton (1997).
At a distance of 2.9 kpc, a proper motion of 1~mas~yr$^{-1}$ corresponds to a transverse velocity of 13.7~km~s$^{-1}$, so that 
the mean vertical motion of the maser features is $-$24.2~km~s$^{-1}$, where the minus sign indicates the direction of the vertical motion away from the Galactic plane (because of the negative galactic latitude $b=-6^\circ .31$ of NGC 281).

If we adopt an estimated distance of 2.2~kpc instead of 2.9~kpc, the resultant mean vertical motion is $-$18.4~km~s$^{-1}$.
If a distance of 3.5 kpc is employed, the mean vertical motion is then calculated similarly to be $-$29.4~km~s$^{-1}$.

If we adopt the IAU standard values $R_0=$8.5~kpc and $\Theta_0=$220~km~s$^{-1}$ (Kerr \& Lynden-Bell 1986) instead of $R_0=$8.0~kpc and $\Theta_0=$200~km~s$^{-1}$ 
and the solar motion relative to the LSR of the velocity $V=$19.5~km~s$^{-1}$ toward $\alpha_{\rm 2000}=$271.$^\circ$0, $\delta_{2000}=$29.$^\circ$0, then the values vary by up to $\sim$10$\%$ in the vertical (DEC) motion.

Our new results from the proper-motion measurements clearly demonstrate that the H$_2$O maser features in NGC~281 West are systematically moving further away from the Galactic plane with a transverse vertical velocity of 20$-$30 km~s$^{-1}$ at its estimated distance of 2.2$-$3.5 kpc.

\subsection{Origin of the Motion Away from the Galactic Plane}

In this section, we consider the following three possible origins of the detected absolute proper motions of the H$_2$O maser features in NGC~281 West, which show a systematic motion away from the Galactic plane as described in the previous section.

First, we look at the possibility that the motions are due to outflows of young stellar objects (YSOs) inducing the maser emission.
Other H$_2$O maser sources show outflow motions of this magnitude (e.g., 10$-$30~km~s$^{-1}$ in W75 reported by Torrelles \etal\ 2003).
Also, the range of radial velocities $V_{\rm LSR}$ of the maser features ($-$39 to $-$22~km~s$^{-1}$) are larger than that of the ambient NGC~281 West cloud ($-$34 to $-$26~km~s$^{-1}$ by Lee \& Jung 2003) or especially of the NW gas clump with maser emission in the cloud ($-$32 to $-$28~km~s$^{-1}$ by Megeath \& Wilson 1997).
It may be therefore suggested that the motions form in outflows.
We regard the internal maser motions within C3 or C1 as possible outflows in each region, and especially the maser features in C3 showed expanding internal motions as we have already seen in $\S$3.1 (figure 2).
However, the internal motions of these maser features in C3 are small ($\sim$10~km~s$^{-1}$; $\sim$0.8$-$0.9~mas~yr$^{-1}$) compared to the absolute proper motion of the H$_2$O maser source, and the central radial velocity of the maser features is in good agreement with that of the entire CO molecular cloud complex of the NGC~281 region (including the NGC~281~West cloud), $V_{\rm LSR}=-$31~km~s$^{-1}$ (Lee \& Jung 2003).
The large separation $\sim$\timeform{8''} between C3 and C1 in the sky, which is equivalent to a projected separation of $\sim$23000~AU ($=$0.11~pc) at an estimated distance of 2.9~kpc, suggests that these two maser components are separately produced by two different YSOs.
On this scale, we only detected the systematic motion in one direction and did not detect bipolar jetlike outflows that are often traced by H$_2$O maser sources.
From these reasons, the detected systematic mean motion of these two components is considered to trace a larger-scale systemic motion of the molecular clouds, rather than motions of outflows.
It is therefore unlikely that the detected systematic motion away from the Galactic plane is due to outflows.  

Second, we take into account the possibility that the motions are due to the rocket effect of HD~5005, which is ionizing the southwestern cloud of NGC~281 (Megeath \& Wilson 1997).
Elmegreen and Lada (1978) suggested that the photoevaporation and expansion of the H\emissiontype{II} region NGC~281 is triggering star formation in the NGC~281 West molecular cloud.
Megeath and Wilson (1997) compared the external pressure of the photoevaporating gas to the internal turbulent pressure of the gas clumps in the NGC~281 West cloud, and found that the shock velocity is low ($\sim$1.5~km~s$^{-1}$).
Also, we can roughly estimate the expansion velocity of the H\emissiontype{II} region from an estimated age of $\sim3$~Myr of HD~5005 (Henning \etal\ 1994) and its location \timeform{3.4'} to the east and \timeform{3.8'} to the north of the NGC~281 West cloud ($\sim$4.3~pc separation at a distance of 2.9~kpc).
Dividing the separation $\sim$\timeform{5'} between NGC~281 West and the H{II} region by the age of HD~5005, 3 Myr, we obtain a rough estimate of the expansion velocity of 0.1~mas~yr$^{-1}$, which has a negligible effect on the observed systemic motion of the NGC~281 region.
Therefore, even though the rocket effect would accelerate the cloud away from the Galactic plane, the effect is unlikely to contribute to the detected motions of the H$_2$O maser features significantly.

Finally, we consider the possibility that the motions are due to the superbubble expansion and blowout.
Megeath \etal\ (2002) and Megeath \etal\ (2003) found a broken ring of the NGC~281 molecular cloud complex $\sim300$~pc above the Galactic plane with a ring diameter of 270~pc, expanding at a line-of-sight velocity of 22~km~s$^{-1}$ parallel to the Galactic plane, which yields a dynamical time of the ring of 6~Myr.
From the velocity-integrated mass in atomic and molecular gas of $3.5\times10^5 M_\odot$ and $10^5 M_\odot$, respectively, they estimated the total kinetic energy of the clouds to be $4.5\times10^{51}$~ergs, which requires multiple supernovae (Megeath \etal\ 2002; Megeath \etal\ 2003).
They suggested that the clouds were created from a fragmenting superbubble shell in a superbubble blowout.
They had no measurement of the motion perpendicular to the Galactic plane.
From our measurements of the systematic motion of the H$_2$O maser features in NGC~281 West away from the Galactic plane, it is suggested that the molecular clouds are moving at a velocity of $20-30$~km~s$^{-1}$ perpendicular to the Galactic plane, which is comparable to the ring expansion velocity of 22~km~s$^{-1}$ parallel to the Galactic plane, thus yielding a comparable kinetic energy ($\sim 4-8\times10^{51}$ ergs) of the clouds for motions perpendicular to the Galactic plane as well as parallel. 
According to the numerical model simulation of fragmenting superbubble blowouts by Mac~Low, McCray and Norman (1989), the fragments can move at velocities $\sim50-100$~km~s$^{-1}$.
The observed velocity of the clouds perpendicular to the Galactic plane, $20-30$~km~s$^{-1}$, is lower than the value expected from the model, but still consistent with the model, taking into account the significant velocity component parallel to the Galactic plane. 

Using the detected vertical velocity $v$ of the region away from the Galactic plane, we can also calculate the timescale of the explosion for the gas in the NGC~281 region to reach the current galactic latitude, assuming that its origin is in the Galactic plane.
On the simplest assumption of a ballistic vertical motion, the dynamical timescale $t$ is calculated from $t=z/v=b/\mu_{Y-{\rm GR}}$, where $z$ is the distance from the Galactic plane, $b$ is the galactic latitude of NGC~281 and $\mu_{Y-{\rm GR}}$ is the derived vertical motion of the maser source in NGC~281 West.
For $b=-$6.$^\circ$31 and $\mu_{Y-{\rm GR}}=-1.77$~mas~yr$^{-1}$, we obtain 13~Myr as a rough estimate of the dynamical timescale of the expansion, which is comparable in the order of magnitude to the dynamical time of 6~Myr by Megeath \etal\ (2002). 
This timescale of $\sim 10$~Myr is consistent with the shorter age of HD~5005, the ionizing source of the H\emissiontype{II} region NGC~281.
According to earlier studies, the upper limit to the lifetime of HD~5005, the driving source of the H\emissiontype{II} region NGC~281, is not much greater than 2~Myr (Elmegreen \& Lada 1978), and the age of HD~5005 is estimated to be $\sim3$~Myr (Henning \etal\ 1994). 
From the difference between the estimated timescale of the superbubble expansion and the age of HD~5005, it is inferred, as suggested by Megeath \etal\ (2002), that the first OB stars including HD~5005 were formed by supernova-driven gas compression after the multiple or sequential supernova explosions
 and that the subsequent and ongoing star formation was triggered afterward by those first-generation OB star cluster as suggested by Elmegreen and Lada (1978).

In conclusion, our measurements of the systemic motion of the H$_2$O maser features in NGC~281 West away from the Galactic plane provide the most direct evidence that the gas in the NGC~281 region and in the HI loop
 was blown out from the Galactic plane, most likely in a superbubble driven by multiple or sequential supernova explosions in the Galactic plane.

\medskip
\noindent {\it Acknowledgements.} We are deeply grateful to the referee Dr.~S.~Thomas~Megeath for his invaluable and detailed comments that significantly improved the paper.  
We sincerely thank Dr.~Philip~Edwards for his careful reading of the manuscript and for very helpful comments and suggestions.
We are grateful to Dr.~Jeremy~Lim for useful information and fruitful discussions at a great opportunity offered by the first Asian Radio Astronomy Winter School in Tokyo, 2007 January.
M.~Sato is with her deepest respect thankful to Dr.~Melanie~Johnston-Hollitt, Prof.~Brian~P.~Schmidt and Prof.~Hiroyuki~Sasada for their kind support and continuous encouragement.

\end{document}